\documentstyle{mn}

\input epsf

\begin{document}
\journal{SUSSEX preprint SUSSEX-AST 98/3-3, astro-ph/9803244}
\title[Galaxy clusters at $0.3<z<0.4$ and the value of $\Omega_{0}$]
{Galaxy clusters at $0.3<z<0.4$ and the value of $\Omega_{0}$}
\author[P.~T.~P.~Viana and A.~R.~Liddle]{Pedro T.~P.~Viana$^{1,2}$ 
and Andrew R.~Liddle$^3$\thanks{Present address: Astrophysics 
Group, The Blackett Laboratory, Imperial College, London SW7 2BZ}\\
$^1$Centro de Astrof\'{\i}sica da Universidade do Porto,
Rua das Estrelas s/c, 4100 Porto, Portugal\\
$^2$Instituto Superior da Maia, Cast\^{e}lo da Maia, 4470 Maia, Portugal\\
$^3$Astronomy Centre, University of Sussex, Falmer, Brighton BN1 9QJ, UK}
\maketitle
\begin{abstract}
The observed evolution of the galaxy cluster X-ray integral temperature
distribution function between $z=0.05$ and $z=0.32$ is used in an attempt to
constrain the value of the density parameter, $\Omega_{0}$, for both open and
spatially-flat universes.  We estimate the overall uncertainty in the
determination of both the observed and the predicted galaxy cluster X-ray
integral temperature distribution functions at $z=0.32$ by carrying out Monte
Carlo simulations, where we take into careful consideration all the most
important sources of possible error.  We include the effect of the formation
epoch on the relation between virial mass and X-ray temperature, improving on
the assumption that clusters form at the observed redshift which leads to an
{\em overestimate} of $\Omega_0$.  We conclude that at present both the
observational data and the theoretical modelling carry sufficiently large
associated uncertainties to prevent an unambiguous determination of
$\Omega_{0}$.  We find that values of $\Omega_{0}$ around 0.75 are most
favoured, with $\Omega_{0}<0.3$ excluded with at least 90 per cent
confidence.  In particular, the $\Omega_{0}=1$ hypothesis is found to be
still viable as far as this dataset is concerned.  As a by-product, we also
use the revised data on the abundance of galaxy clusters at $z=0.05$ to
update the constraint on $\sigma_8$ given by Viana \& Liddle \shortcite{VL},
finding slightly lower values than before.
\end{abstract}

\section{Introduction}

The number density of rich clusters of galaxies at the present epoch has been 
used to constrain the amplitude of mass density fluctuations on a 
scale of $8\,h^{-1}\,{\rm Mpc}$ (Evrard 1989; Henry \& Arnaud 1991; 
White, Efstathiou \& Frenk 1993a; Viana \& Liddle 1996, henceforth VL; 
Eke, Cole \& Frenk 1996; Kitayama \& Suto 1997). This is usually referred to 
as $\sigma_{8}$, where $h$ is the present value of the Hubble parameter, 
$H_{0}$, in units of $100\;{\rm km}\,{\rm s}^{-1}\,{\rm Mpc}^{-1}$. 
However, the derived value of $\sigma_{8}$ depends to a great extent on 
the present matter density in the Universe, parameterized by $\Omega_{0}$, 
and more weakly on the presence of a cosmological constant, $\Lambda$. 
The cleanest way of breaking this degeneracy is to include information 
on the change in the number density of rich galaxy clusters with 
redshift \cite{FWED}, the use of X-ray clusters for this purpose having been 
proposed by Oukbir \& Blanchard \shortcite{OB} and subsequently further 
investigated \cite{HM,OB97}. Several attempts have been made 
recently, with wildly differing results 
\cite{Henry,FBC,Grossetal,BB,Ekeetal,Retal}.

The best method to find clusters of galaxies is through their X-ray emission,
which is much less prone to projection effects than optical identification.
Further, the X-ray temperature of a galaxy cluster is at present the most
reliable estimator of its virial mass.  This can then be used to relate the
cluster mass function at different redshifts, calculated for example within
the Press--Schechter framework \cite{PS,BCEK}, to the observed cluster X-ray
temperature function. We can therefore compare the evolution in the number
density of galaxy clusters seen in the data with the theoretical expectation
for large-scale structure formation models, which depends significantly only 
on the assumed
values of $\Omega_{0}$ and $\lambda_{0}\equiv\Lambda/3H^{2}_{0}$, the latter 
being the contribution of $\Lambda$ to the total present energy density in 
the Universe.

However, the X-ray temperature of a cluster of galaxies is not an easily 
measurable quantity, as compared to its X-ray luminosity.  A minimum flux
is required, so that there is a high enough number of photons for the
statistical errors in the temperature determination to be reasonably small.
Because of this, although estimates of the present-day cluster X-ray
temperature function have been available since the early 90's \cite{ESFA,HA},
the change in the cluster X-ray temperature function as we look further into 
the
past has been much more difficult to determine.  Estimates for the X-ray
temperatures of individual clusters with redshifts as high as 0.3 have been
available for some years (e.g.~see David et al.~1993), but only with the
advent of the {\em ASCA} satellite has it been possible to measure X-ray
temperatures for clusters of galaxies around that redshift in a systematic
way, and to go to even higher redshifts.

The evolution of the cluster X-ray luminosity function with redshift, though
easier to determine, provides much weaker constraints on $\Omega_{0}$ and
$\lambda_{0}$, due to the fact that the X-ray luminosity of a galaxy cluster
is not expected to be a reliable estimator of its virial mass (e.g.~Hanami
1993). Though it could in principle provide some indication of the change 
of the cluster X-ray temperature function with redshift, the problem is that 
not only is there considerable scatter in the present-day cluster 
X-ray temperature verses luminosity relation \cite{Davetal,Fetal}, but 
it is also not known how the relation may change with redshift, though 
recently it has been argued that at least up to $z=0.4$ it does not seem to 
evolve \cite{MScharf,AF,Retal,SBO}. 

The deepest complete X-ray sample of galaxy clusters presently available is
the one obtained from the {\em Einstein Medium Sensitivity Survey} ({\em
EMSS}) \cite{Getal,Hetal}.  This sample is restricted to objects with
declination larger than $-40^{\rm o}$ and is flux-limited, with $F_{{\rm
obs}}\geq1.33\times10^{-13}\;{\rm erg}\,{\rm cm}^{-2}\,{\rm s}^{-1}$, where
$F_{{\rm obs}}$ is the cluster flux in the 0.3 to 3.5 keV band which falls in
a $2'.4\times2'.4$ {\em EMSS} detect cell.  It presently contains 90 clusters
of galaxies, after a few misidentifications were recently removed
\cite{GioiaL,Netal}.  This is the only complete galaxy cluster catalogue 
beyond
a redshift of 0.3, and as such unique in providing the means to distinguish
between different possible values for $\Omega_{0}$ and $\lambda_{0}$.
However, until the recent effort by Henry \shortcite{Henry}, very few X-ray
temperatures were known for those galaxy clusters in the {\em EMSS} sample
with redshifts exceeding 0.15 (see Sadat et al.~1998 for a recent
compilation).  Henry \shortcite{Henry} used {\em ASCA} to observe all galaxy
clusters in the {\em EMSS} cluster sample with $0.3\leq z \leq 0.4$ and
$F_{{\rm obs}}\geq2.5\times10^{-13}\;{\rm erg}\,{\rm cm}^{-2}\,{\rm s}^{-1}$.
The resulting sub-sample of 10 clusters has a median redshift of 0.32, and
the data obtained for each cluster, the X-ray flux, luminosity and
temperature, can be found in Table 1 of Henry (1997).

We will use this data together with the present-day (median redshift 0.05)
cluster X-ray temperature function.
We work within the extended Press--Schechter formalism proposed by Lacey
\& Cole (1993, 1994), which allows an estimation of the formation times of
dark matter halos.  We will assume the dark matter to be cold, and consider
the cases of an open universe, where the cosmological constant is zero, and a
spatially-flat universe, such that $\lambda_{0}=1-\Omega_{0}$.

\section{The theoretical mass and temperature functions}

The usual means by which the mass function of virialized structures can be
determined analytically is through the Press--Schechter approach, which has
been found to reproduce well the mass functions recovered from various
$N$-body simulations \cite{LC,ECF,T}.  The comoving number density of galaxy
clusters in a mass interval ${\rm d}M$ about $M$ at a redshift $z$ is 
given by
\begin{eqnarray}
\label{mfa}
n(M,z)\,{\rm d}M= & & \\ \nonumber
& & \hspace*{-2cm} -\sqrt{\frac{2}{\pi}}\frac{\rho_{\rm b}}{M}\,
\frac{\delta_{\rm c}}{\sigma^2(R,z)}\frac{{\rm d}\sigma(R,z)}
{{\rm d}M}\exp\left[-\frac{\delta_{\rm c}^2}{2\sigma^2(R,z)}\right]{\rm 
d}M\,,
\end{eqnarray}
where $\rho_{\rm b}$ is the comoving matter density, $\sigma(R,z)$ is the
dispersion of the linearly-evolved density field smoothed by a top-hat
window function on the comoving scale $R$, such that $R^3=3M/4\pi\rho_{\rm
b}$, and $\delta_{\rm c}$ is the linear density threshold
associated with the collapse and subsequent virialization of the galaxy
clusters.  This last quantity depends to some extent on the geometry of the
collapsing structures \cite{Mon}, but since rich galaxy clusters are
relatively rare, it would seem to be a fair assumption to take their
collapse to be close to spherical \cite{Ber}.  The analytical calculation 
would then yield $\delta_{\rm c}=1.69$ in the case of an Einstein--de Sitter
universe, with a decrease at most of $5$ per cent when one goes to a universe
with sub-critical density, as long as $\Omega_{0}\geq0.1$, whether or not a
cosmological constant is present \cite{ECF}.  This is supported by the
results from $N$-body simulations, which prefer $\delta_{\rm c} = 1.7\pm0.2$
\cite{LC,ECF,T}.  As we want to be conservative, we will allow for this 
margin
of variation in the value of $\delta_{\rm c}$, assuming it to be equivalent
to a 95 per cent confidence interval.

If we have in mind a particular shape for the power spectrum of density 
perturbations, we can further simplify equation (\ref{mfa}) by writing 
the derivative in terms of $\sigma(R,z)$. As we will be assuming all the dark 
matter to be cold, the value of $\sigma(R,z)$ in the vicinity of $8h^{-1}$ 
Mpc can be obtained to a good approximation through,
\begin{equation}
\label{sig}
\sigma(R,z)=\sigma_{8}(z)\left(\frac{R}{8 h^{-1}\;{\rm 
Mpc}}\right)^{-\gamma(R)}\,,
\end{equation}
where 
\begin{equation}
\label{del}
\gamma(R)=(0.3\Gamma+0.2)\left[2.92+\log_{10}
        \left(\frac{R}{8 h^{-1}\;{\rm Mpc}}\right)\right]\,,
\end{equation} 
and $\Gamma$ is the usual shape parameter of the cold dark matter (CDM) 
transfer function.

Note that $\gamma(R)$ is independent of $z$, reflecting the fact
that the shape of the power spectrum of density perturbations in the case of
CDM models does not change after the epoch of matter--radiation equality.  We
will assume that $\Gamma=0.230^{+0.042}_{-0.034}$ at the 95 per cent
confidence level, for which a good fit to the observed present shape of the
galaxy and cluster correlation functions in the vicinity of $8h^{-1}$ Mpc is 
obtained \cite{PD,VL}. It should be mentioned however that when the abundance 
of X-ray clusters over a range of temperatures is used, the preferred value 
for $\Gamma$ seems to be closer to 0.10 \cite{ECF,Ekeetal}. Though in this 
case $\Gamma$ is not as well constrained, there is clearly a significant 
disagreement between the values obtained through the two methods. The 
source of this discrepancy needs to be investigated, and may lie ultimately 
in a failure of models whose power spectrum can be parameterized through 
$\Gamma$ to reproduce the observed perturbation spectrum (see e.g. Peacock 
1997). 

The quantity $\sigma_{8}(z)$ is related with its present value 
$\sigma_{8}(0)$ via the perturbation growth law 
\begin{equation}
\label{growth}
\sigma_8(z) = \sigma_8(0) \, \frac{g(\Omega(z))}{g(\Omega_0)} \, 
\frac{1}{1+z} \,,
\end{equation}
where the appropriate formulae for $g(\Omega)$ and $\Omega(z)$, depending on
whether the universe is open or spatially-flat, can be found in VL 
[equations (8) and (10), and (9) and (11), respectively]. 

Using expression (\ref{sig}) we can now calculate the derivative appearing in 
equation (\ref{mfa}), and substituting we end up with
\begin{eqnarray}
\label{mf}
n(M,z) \, {\rm d}M = \sqrt{\frac{2}{\pi}}\frac{\rho_{\rm b}}{M^2} 
\frac{\delta_{\rm c}}{3\sigma(R,z)}(0.3\Gamma+0.2)\times & & \\ \nonumber
& & \hspace*{-6.5cm} \left[2.92+\log_{10}
\left(\frac{R}{8 h^{-1}\;{\rm Mpc}}\right)\right]\exp
\left[-\frac{\delta_{\rm c}^{2}}{2\sigma^{2}(R,z)}\right]{\rm d}M.
\end{eqnarray}

In order to transform this mass function into the cluster X-ray temperature 
function, one needs to relate the virial mass of a galaxy cluster to its 
X-ray temperature. Such a relation has been obtained analytically by 
Lilje \shortcite{lilje} (see also Hanami 1993), and confirmed through 
hydrodynamical $N$-body simulations \cite{NFW,BN}, 
\begin{eqnarray}
\label{mvprop}
M_{\rm v} \propto \Omega_{0}^{-1/2}\,\chi^{-1/2}\,
\left(2\frac{r_{\rm v}}{r_{{\rm m}}}\right)^{3/2} \times & & \\ \nonumber
& & \hspace*{-4.0 cm}\left[1-\eta\left(\frac{r_{\rm v}}{r_{\rm m}}
\right)^{3}\right]^{-3/2}\,(1+z_{\rm m})^{-3/2}\,(k_{\rm B}T)^{3/2}h^{-1}\,,
\end{eqnarray}
where $z_{\rm m}$ is the redshift of cluster turnaround, and 
\begin{equation}
\chi = \left(\frac{4}{3\pi}\right)^{2}\xi\,,
\end{equation}
\begin{equation}
\eta = 2\left(\frac{4}{3\pi}\right)^{2}\left(\frac{\lambda_{0}}{\Omega_{0}}
\right)\chi^{-1}(1+z_{\rm m})^{-3}\,,
\end{equation}
with $\xi$ the ratio between the cluster and background densities at 
turnaround. This quantity was calculated numerically in VL, 
depending only on $\Omega\equiv\Omega(z_{\rm m})$ via 
$\chi=\Omega^{-b(\Omega)}$, where $b(\Omega)=0.76-0.25\,\Omega$ in the case 
of an open universe, and $b(\Omega)=0.73-0.23\,\Omega$ for a spatially-flat 
universe.

The radii of turnaround and virialization, respectively 
$r_{\rm m}$ and $r_{\rm v}$, are related through 
\begin{equation}
\frac{r_{\rm v}}{r_{\rm m}} = \frac{1-\eta/2}{2-\eta/2}\,,
\end{equation}
in the case when a galaxy cluster is assumed to be an ideal virialized 
system collapsed from a top-hat perturbation. The presence of significant 
substructure during collapse would lead to dynamical relaxation thus 
making the clusters more compact, decreasing the ratio 
$r_{\rm v}/{r_{\rm m}}$. 

The proportionality constant in expression (\ref{mvprop}) can be obtained
either through analytical derivation assuming hydrostatic equilibrium (e.g.
see Bryan \& Norman 1998), or by using results from hydrodynamical $N$-body
simulations.  We choose the latter option, as it provides an estimate of the
uncertainties involved, allowing for the possibility of deviations from
hydrostatic equilibrium due for example to bulk gas motions and turbulence
\cite{NB}.  The hydrodynamical $N$-body simulations we use are those of White
et al.~\shortcite{WNEF}, carried out in an Einstein--de Sitter universe.
They imply that a galaxy cluster with a X-ray temperature of 7.5 keV has a
virial mass of $M_{\rm v}=(1.23\pm0.32)\times10^{15}\,h^{-1}\;{\rm
M}_{\sun}$, for an estimated virialization redshift of $z_{\rm
c}\simeq0.05\pm 0.05$ \cite{ME,NFW}.  This corresponds to a turnaround
redshift of $z_{\rm m}\simeq0.67\pm 0.08$, since $z_{\rm c}$ and $z_{\rm
m}$ are related by the fact that the time of collapse and subsequent
virialization, $t_{\rm c}$, is twice the time of maximum expansion, $t_{\rm
m}$, and in an Einstein-de Sitter universe
\begin{equation}
t = \frac{2}{3}H_{0}^{-1}(1+z)^{-3/2}\,.
\end{equation}
Putting all this together we are now able to normalize expression 
(\ref{mvprop}),
\begin{eqnarray}
\label{mv}
M_{\rm v}=(1.23 \pm 0.33)\times 10^{15}\,\Omega_{0}^{-1/2}\,
        \chi^{-1/2}\,\left(2 \frac{r_{\rm v}}{r_{\rm m}} 
        \right)^{3/2} \times & & \\ \nonumber
 & & \hspace*{-8 cm}\left[1-\eta 
        \left(\frac{r_{\rm v}}{r_{\rm m}}\right)^{3}\right]^{-3/2}
        \left(\frac{1.67}{1+z_{\rm m}}\right)^{3/2}
        \left(\frac{k_{\rm B}T}{7.5\,{\rm keV}}\right)^{3/2}h^{-1}\;
        {\rm M}_{\sun}\,,
\end{eqnarray}
where the error is 1-sigma. This normalization of the virial mass---X-ray 
temperature relation agrees very well with that obtained by Bryan \& 
Norman \shortcite{BN}, who used the largest set of hydrodynamical 
$N$-body simulations ever assembled.

We now have the problem that even after the background cosmology is chosen, 
by fixing $\Omega_{0}$ and $\lambda_{0}$, expression (\ref{mv}) depends on 
the
redshift of cluster turnaround, $z_{\rm m}$.  As this can be determined from
the virialization redshift, $z_{\rm c}$, using the fact that $t_{\rm
m}=2t_{\rm c}$ [for expressions of $t$ as a function of $z$ in open and
spatially-flat universes see VL, equations (26) and (27)], we simply
need a means to estimate the distribution of the redshifts of cluster
virialization at each given virial mass.  The most well justified method
which provides this was put forward by Lacey \& Cole (1993,
1994), though it may slightly underestimate $z_{\rm c}$ \cite{T}.

Lacey \& Cole constructed a merging history for dark matter halos based 
on the random walk trajectories technique, and derived an analytical 
expression for the probability that a galaxy cluster with present virial 
mass $M$ would have formed at some redshift $z$, 
\begin{equation}
\label{pzlc}
p(z) = p(w(z)) \frac{{\rm d}w(z)}{{\rm d}z}\,,
\end{equation}
where
\begin{eqnarray}
\label{pwlc}
p(w(z)) = 2 \, w(z) \left( f^{-1}-1 \right) \, {\rm erfc} 
        \left(\frac{w(z)}{\sqrt{2}}\right) - & & \\ \nonumber
& & \hspace*{-5.0 cm} \sqrt{\frac{2}{\pi}} 
        \left( f^{-1}-2 \right) \exp{\left(-\frac{w^{2}(z)}{2}\right)} \,,
\end{eqnarray} 
and
\begin{equation}
w(z) = \frac{\delta_{{\rm c}}\left(\sigma(M,0)/\sigma(M,z) - 1 \right)}
        {\sqrt{\sigma^2 (fM,0) - \sigma^2 (M,0)}}\,,
\end{equation}
with $f$ the fraction of the cluster mass assembled by redshift $z$.
Independently of background cosmology we will assume $f=0.75\pm0.15$, as
after this mass fraction has been assembled it is expected that the X-ray
temperature of a cluster of galaxies will not change significantly
\cite{NFW}.  We will consider the uncertainty in the value of $f$ to roughly
correspond to a 95 per cent confidence interval.  Although the expression for
the formation probability given above was obtained for a power-law spectrum
of matter density fluctuations with index $n=0$, while at the cluster scale
$n$ is expected to be close to $-2$ \cite{Hetal,H,OBB,Mark98}, numerical
results show that $p(w(z))$ depends only very weakly on $n$ (Lacey \& Cole
1993).

The present comoving number density of galaxy clusters 
per temperature interval ${\rm d}(k_{\rm B}T)$ with a mean X-ray 
temperature of $k_{\rm B}T$, that formed at each redshift $z$, can 
now be calculated using the chain rule and equation (\ref{mf}) 
with $z=0$,
\begin{equation}
\label{lcrn}
n(k_{\rm B}T,z)\,{\rm d}(k_{{\rm B}}T)\,{\rm d}z = 
        \frac{3}{2}\frac{M}{k_{{\rm B}}T}\,n(M,z=0) \, p(z)
        \, {\rm d}M\,{\rm d}z \,.
\end{equation}
The present cluster X-ray temperature function at $k_{\rm B}T$ is obtained by 
integrating this expression from redshift zero up to infinity, with the 
virial mass $M$ obtained through expression (\ref{mv}) taking into account 
the assumed $k_{\rm B}T$ and the value of the integration variable $z=z_{\rm 
c}$. 

The cluster X-ray temperature function at any redshift $z$, 
for some temperature $k_{\rm B}T$, can be obtained by taking the point of 
view of someone placed at that redshift, i.e. transferring the conditions 
prevalent at that redshift to the present epoch. For example, one takes 
$\Omega_{0}=\Omega({\rm z})$, and changes the normalization of expression 
(\ref{mv}) so that for $z=0$ one obtains the virial mass---X-ray temperature 
relation that applies at the redshift of interest. 

In order to compare with the available data, we will actually need to
calculate the cumulative or integral cluster X-ray temperature function,
$N(>k_{\rm B}T, z)$, i.e.  the comoving number density of galaxy clusters
with an X-ray temperature exceeding $k_{\rm B}T$ at redshift $z$.  This is
obtained from the differential cluster X-ray temperature function, $n(k_{\rm
B}T, z){\rm d}(k_{\rm B}T)$, by integrating it from the minimum X-ray
temperature required up to infinity.

\section{Comparing observational data with theoretical expectations}

We use two pieces of data, the integral cluster X-ray temperature 
functions at $z=0.05$ and $z=0.32$, where these are the median redshifts of 
the X-ray cluster samples we are considering. 

\subsection{The low-redshift data}

The integral cluster X-ray temperature function at $z=0.05$ can be determined 
using 
the dataset presented in Henry \& Arnaud \shortcite{HA}. We actually 
use an updated version of it, with more accurate X-ray flux and temperature 
determinations, which has been kindly provided to us by Pat Henry. Without 
taking into consideration the temperature measurement errors, we derived for 
the 
number density of galaxy clusters at $z=0.05$ with X-ray temperature 
exceeding 
6.2 keV
\begin{equation}
\label{nd}
N(>6.2\,{\rm keV},\,0.05)=1.80\times10^{-7}h^{3}
\;{\rm Mpc}^{-3}\,,
\end{equation}
which as expected agrees with the value obtained by Eke et al.  (1996,1998)
and Henry (1997).  It is also in good agreement with the results of Edge
et al.~\shortcite{ESFA} and Markevitch \shortcite{Mark98}.  

However, what not
many people take into account is that the existence of temperature 
measurement 
errors makes the above value a biased estimator of the real value of 
$N(>6.2\,{\rm keV},\,0.05)$ in the Universe.\footnote{We are very much 
indebted to Alain Blanchard for pointing this out to us.}  This can be easily 
seen by constructing a large number of datasets with the same clusters in 
each 
as in the $z=0.05$ dataset, but where the X-ray temperature for each cluster 
is randomly drawn from a Gaussian distribution with mean and dispersion as
observed for that cluster.  This procedure simulates the repetition of the
temperature measurements, assuming the cluster X-ray temperatures originally
measured are the actual ones.  If one now determines $N(>6.2\,{\rm
keV},\,0.05)$ for each cluster dataset thus obtained, the mean of the
distribution turns out to be $2.12\times10^{-7}h^{3}\;{\rm Mpc}^{-3}$.  
Hence, the existence of measurement errors in the X-ray temperature 
determinations leads to an {\em overestimation} of the 
real value for $N(>6.2\,{\rm keV},\,0.05)$ if one uses the 
actual measured temperatures.  This can be easily understood
if one remembers that as the X-ray temperature goes up, the cluster number
density decreases.  Given that to a first approximation it is as probable for
a cluster with actual X-ray temperature below 6.2 keV to have a measured
temperature above that value, as it is for a cluster with $k_{\rm B}T>6.2$ 
keV
to have a smaller measured temperature, the net effect will therefore be an
apparent increase in the number density of galaxy clusters with X-ray
temperature above 6.2 keV.  We expect $1.80\times10^{-7}h^{3}\;{\rm
Mpc}^{-3}$ to be an overestimation of the real value for $N(>6.2\,{\rm
keV},\,0.05)$ in the same proportion as $2.12\times10^{-7}h^{3}\;{\rm
Mpc}^{-3}$ exceeds the assumed correct value
$1.80\times10^{-7}h^{3}\;{\rm Mpc}^{-3}$ used in constructing the artificial
datasets above.  Therefore the corrected best estimate for the number density
of galaxy clusters at $z=0.05$ with X-ray temperature exceeding 6.2 keV is
\begin{equation}
\label{ndlzobs}
N(>6.2\,{\rm keV},\,0.05)=1.53\times10^{\pm0.16}\times10^{-7}h^{3}
\;{\rm Mpc}^{-3}\,,
\end{equation}
where the errors represent 1-sigma confidence levels.  They were obtained
through a bootstrap procedure, which allows an estimation of the uncertainty
associated with the sampling variance, where we constructed $10^{4}$
cluster datasets by randomly selecting, with replacement, from the original
list of 25 X-ray clusters in Henry \& Arnaud \shortcite{HA}.  The number of
clusters in each sample is drawn from a Poisson distribution with mean 25, in
order to include the counting error in the uncertainty.  Each time a cluster
is selected its X-ray temperature is estimated by randomly drawing from a
Gaussian distribution with the mean and dispersion observed for the cluster.
In this way the temperature measurement errors also contribute to the total
uncertainty.

In Eke et al.~(1998), that X-ray temperature measurement errors lead
to an overestimate of the real value for $N(>k_{\rm B}T,\,z)$ is dealt
with through a Gaussian smoothing of the temperature distribution
function predicted in each $(\Omega_{0},\lambda_{0})$ cosmology, which
is then integrated to give $N(>k_{\rm B}T,\,z)$.

There are several reasons why we chose to concentrate on galaxy clusters with 
X-ray temperature exceeding 6.2 keV. The first is that $N(>6.2\,{\rm keV})$ 
best represents the mean curve going through the observational 
points for $N(>k_{\rm B}T)$, both at $z=0.05$ and $z=0.32$, as can be 
seen in Figure 2 of Henry \shortcite{Henry}.  Also, the Press--Schechter 
framework should work best on the largest scales, i.e. for the 
highest masses and X-ray temperatures, as in hierarchical
cosmologies these are the ones for which the density field has evolved least,
therefore keeping its gaussianity to a greater extent.  Related to this is
the problem of shear, which starts becoming an important factor in the 
collapse of density perturbations as the density field develops, leading
to deviations from the idealized spherical collapse situation.  Another
reason is that in the normalization of the relation between virial mass and
X-ray temperature for galaxy clusters, we used hydrodynamical $N$-body
simulations which do not take into account a possible (pre-)heating of the
intracluster medium due to starbursts and supernovae in the galaxies.  This
effect is still quite difficult to model realistically, but the few attempts
that have been made seem to show that it becomes more important as the
cluster virial mass decreases.  For a galaxy cluster whose X-ray temperature
would otherwise be 5 keV, the heating may increase the cluster X-ray
temperature by as much as 15 per cent \cite{ME,NFW}. 

\subsection{The high-redshift data}

The comoving number density of galaxy clusters with X-ray temperature
exceeding 6.2 keV at $z=0.32$ can be calculated using the {\em EMSS}
sub-sample of 10 clusters with redshifts between 0.3 and 0.4 and fluxes above
$2.5\times10^{-13}\;{\rm erg}\,{\rm cm}^{-2}\,{\rm s}^{-1}$, for which Henry
\shortcite{Henry} obtained the mean X-ray temperatures through {\em ASCA}.
We used the data in Table 1 of Henry \shortcite{Henry}, regarding the X-ray
flux and temperature for each cluster, to estimate the integral cluster X-ray
temperature function at $z=0.32$.  We did this both for open
and spatially-flat universes, using the estimator
\begin{equation}
\label{est}
N(>k_{{\rm B}}T,\,z)=\sum_{i=1}\frac{1}{V_{{\rm max},i}}\,,
\end{equation}
where the sum is over all clusters with $k_{{\rm B}} T_{i}>k_{{\rm
B}}T$, and $V_{{\rm max},i}$ is the maximum volume in which cluster
$i$ could have been detected at the $4\sigma$ level in the {\em EMSS}
within the redshift shell under consideration (0.3 to 0.4 in our
case).

The steps which need to be taken in order to calculate these volumes are
described in detail in Henry et al.~\shortcite{Hetal}.  They involve the
determination of the maximum redshift at which each galaxy cluster could have
been detected as a function of its observed flux, using equations (1) and (2)
of Henry et al.~\shortcite{Hetal}.  In this calculation one has to
compensate for the fact that clusters of galaxies are extended objects, and
thus some of their flux will be outside the {\em EMSS} detect cell.  One
therefore needs to estimate the typical core radius, from which most of the
flux originates, of the galaxy clusters in the {\em EMSS} sub-sample from
Henry \shortcite{Henry}.  In the absence of data specific to this sub-sample,
we use the ratio between the extended and detect cell fluxes estimated in
Henry et al.~\shortcite{Hetal} for a sample of 4 galaxy clusters extended
within the {\em EMSS} with a mean redshift of 0.29.  They find this ratio to
be equal to $2.10\pm0.19$, where we will assume the error to be $1\sigma$.
This implies a typical core radius around $0.15\,h^{-1}\;{\rm Mpc}$, 
depending on the chosen values for $\Omega_{0}$ and $\lambda_{0}$, which 
we will assume to remain the same for $0.3<z<0.4$.

The detection volume for a given galaxy cluster is then obtained by summing, 
over all limiting fluxes of the {\em EMSS} (see Table 3 of Henry et
al.~1992) starting at $2.5\times10^{-13}\;{\rm erg}\,{\rm cm}^{-2}\,{\rm
s}^{-1}$, the volumes lying between a redshift of 0.3 and whichever is the
lesser of 0.4 and the maximum detection redshift for the cluster. The errors
affecting the calculation of the detection volumes are thus those associated
with the flux measured for each galaxy cluster and the compensation for the
extended nature of galaxy clusters.

As in the lower redshift case, the calculation of $N(>6.2\,{\rm
keV},\,0.32)$ using the X-ray temperatures measured for the galaxy
clusters found between $z=0.3$ and $z=0.4$ would lead to the
overestimation of $N(>6.2\,{\rm keV},\,0.32)$ due to the presence of
errors in the X-ray temperature determinations.  Again, we correct for
this by simulating the repetition of the temperature measurements.
The ratio between the mean value obtained for $N(>6.2\,{\rm
keV},\,0.32)$ from all the simulated datasets, and the value one gets
for $N(>6.2\,{\rm keV},\,0.32)$ using the original dataset, provides an
estimate for the expected ratio between the latter and the real value
for $N(>6.2\,{\rm keV},\,0.32)$ in the Universe.  After performing
this correction, the best estimate for the number density of galaxy
clusters at $z=0.32$ with X-ray temperature exceeding 6.2 keV becomes
\begin{equation}
\label{ndhzobs}
N(>6.2\,{\rm keV},\,0.32)=3.98\,\Omega_{0}^{B(\Omega_{0})}\times10^{-8}
h^{3}\;{\rm Mpc}^{-3}\,,
\end{equation}
where $B(\Omega_{0})=0.09+0.38\,\Omega_{0}-0.29\,\Omega_{0}^{2}$ if the
Universe is open and
$B(\Omega_{0})=0.25+0.94\,\Omega_{0}-0.78\,\Omega_{0}^{2}$ if the Universe is
spatially--flat. These fitting functions have an associated error of less
than one per cent and are valid for $\Omega_{0}$ between 0.1 and 1.

The bootstrap method is the simplest way of simulating the procedure 
involved in the extraction of a sample from a given distribution. 
In our case this distribution is simply the population of galaxy clusters 
in the Universe in the redshift bin $0.3<z<0.4$ with {\em EMSS} X-ray fluxes 
exceeding $2.5\times10^{-13}\;{\rm erg}\,{\rm cm}^{-2}\,{\rm s}^{-1}$. The 
bootstrap method therefore allows the estimation of the {\em dispersion} one 
would expect to obtain in the values measured for some quantity related to 
that population, for example $N(>6.2\,{\rm keV},\,0.32)$, if the
type of sampling that led to the dataset in Henry \shortcite{Henry} was
repeated a large number of times across the sky. 

\subsection{The method of comparison}

Let us now assume that in our Universe $N(>6.2\,{\rm keV},\,0.32)$
takes some particular overall value. We would then expect this value
to be the {\em mean} of the distribution function assembled with the
values that would be measured for $N(>6.2\,{\rm keV},\,0.32)$ if the
type of sampling that led to the dataset in Henry \shortcite{Henry}
was repeated a large number of times across the sky.  On the other
hand, we would expect that the {\em shape} of this distribution would
be that obtained through the bootstrap procedure mentioned at the end
of the previous subsection.  We are therefore now in a position to ask
the following question. If $N(>6.2\,{\rm keV},\,0.32)$ took some
overall value in the Universe, how probable would it be to measure the
value for $N(>6.2\,{\rm keV},\,0.32)$ given by the dataset in Henry
\shortcite{Henry}, after correcting it for the displacement due to
errors in the X-ray temperature measurements? We can then attach, for
each value of $\Omega_{0}$, a probability of the value for
$N(>6.2\,{\rm keV},\,0.32)$ given by the dataset in Henry
\shortcite{Henry} being actually measured. The {\em exclusion level}
on each value of $\Omega_{0}$ is obtained simply by subtracting
this probability from one.

In summary the following steps were taken, so that an exclusion level
can be associated with each $\Omega_0$ based on the X-ray cluster
datasets for $z=0.05$ and $z=0.32$:
\begin{enumerate}
\item The Universe was assumed to be either open or spatially-flat, 
with $\Omega_{0}$ taking a value between 0.1 and 1.
\item The best estimate for $N(>6.2\,{\rm keV},\,0.32)$ in the Universe, 
given the dataset in Henry~\shortcite{Henry}, was calculated. This is the 
result shown in equation (\ref{ndhzobs}), after correcting for the effect 
of the X-ray temperature measurement errors. 
\item A bootstrap procedure analogous to that described for 
the $z=0.05$ data was performed in order to determine the expected 
{\em shape} for the distribution function of $N(>6.2\,{\rm keV},\,0.32)$,
if the type of sampling that led to the dataset in Henry \shortcite{Henry} 
was repeated a large number of times across the sky. The number of clusters 
in each sample is now drawn from a Poisson distribution with mean 10, and  
the input observational errors (in the typical ratio of extended to {\em 
EMSS} detect cell fluxes at $z=0.29$, and in the {\em ASCA} X-ray fluxes and 
temperatures) are modelled as Gaussian distributed. 
\item Using the method described in the section 2, the 
theoretically-expected overall value for $N(>6.2\,{\rm keV},\,0.32)$ 
given the assumed $\Omega_{0}$ was calculated. The normalization
$\sigma_8$ of the spectrum was fixed by the low-redshift data using
equation (\ref{ndlzobs}).
\item The distribution function for $N(>6.2\,{\rm keV},\,0.32)$, determined 
through the bootstrap procedure, was modified by dividing the values obtained 
for $N(>6.2\,{\rm keV},\,0.32)$ by their mean and multiplying them by the 
value determined in (iv), so that this value becomes the new mean and the 
relative shape of the distribution is maintained. 
\item We calculated the probability of obtaining a value as high, or as low, 
as that determined in (ii), given the distribution constructed in (v). The 
exclusion level on the assumed $\Omega_{0}$ equals one minus this 
probability.  
\end{enumerate}

It should be noted that in step (iv), the calculation of the
theoretically-expected overall value for $N(>6.2\,{\rm keV},\,0.32)$
for each assumed $\Omega_{0}$ has some theoretical uncertainty
associated with it.  This results from the uncertainties associated
with the values one should use for the normalization of the cluster
X-ray temperature to virial mass relation, $\delta_{{\rm c}}$, $f$ and
$\Gamma$. There is also the uncertainty associated with the observed
value for $N(>6.2\,{\rm keV}, 0.05)$. The total uncertainty in the
theoretically-expected overall value for $N(>6.2\,{\rm keV},\,0.32)$,
for each assumed $\Omega_{0}$, will be calculated through Monte Carlo
simulations described in section 4, which will also explain the
procedure used to incorporate this uncertainty into the calculation of
the final exclusion level on each assumed $\Omega_{0}$.

In general, it is incorrect to calculate the exclusion level
associated with some theoretically-expected overall value by directly
using the distribution obtained through a bootstrap procedure, and
asking how probable it would be to obtain such a high (or low) value
given that distribution.  In practice, however, if the bootstrap
distribution is symmetric, the exclusion level calculated this way,
and in the more correct way described above in points (v) and (vi), is
nearly the same.  Nevertheless, in the case of the high-redshift
cluster data, the distribution for $N(>6.2\,{\rm keV},\,0.32)$
obtained through the bootstrap procedure is highly asymmetric with a
long right-sided tail.  Given that high $\Omega_{0}$ models tend to
predict an overall value for $N(>6.2\,{\rm keV},\,0.32)$ which is
smaller than the value determined through the dataset in Henry
\shortcite{Henry}, identifying the mean of the bootstrap distribution
with the latter rather than with the former would lead to an
exaggerated difficulty for higher $\Omega_{0}$ models to reproduce the
observations, and thus to a higher exclusion level.  On the contrary,
lower $\Omega_{0}$ models, which tend to predict the inverse, would
benefit from incorrectly calculating the exclusion level.  In the next
section we will estimate this effect.

In the case of the low-redshift data, though the bootstrap procedure yields a
distribution for $N(>6.2\,{\rm keV},\,0.05)$ which is best characterized as a
lognormal, in practice the deviation from symmetry is sufficiently small for
it to be preferable, given the much simpler calculations involved, to
associate the bootstrap uncertainty with the corrected value for
$N(>6.2\,{\rm keV},\,0.05)$ determined from the dataset in Henry \& Arnaud
\shortcite{HA}, rather than with some theoretically-expected overall value.

In Figure 1 we show the binned probability distributions obtained through the
bootstrap method for \mbox{$N(>6.2\,{\rm keV},\,0.32)$}, in the cases of
$\Omega_{0}=1$ and 0.3, where the Universe is assumed either open or
spatially-flat.  The distributions were altered so that their mean now
coincides with the theoretically-expected overall value in those cases.  It 
also
shows the corrected value for $N(>6.2\,{\rm keV},\,0.32)$ determined from the
dataset in Henry \shortcite{H}.  The peaks in the distributions are not an
artifact of the chosen binning, but correspond to different numbers of galaxy
clusters in each bootstrap sample having an X-ray temperature in excess of
our chosen threshold 6.2 keV.

\begin{figure}
\centering
\leavevmode\epsfysize=5.4cm \epsfbox{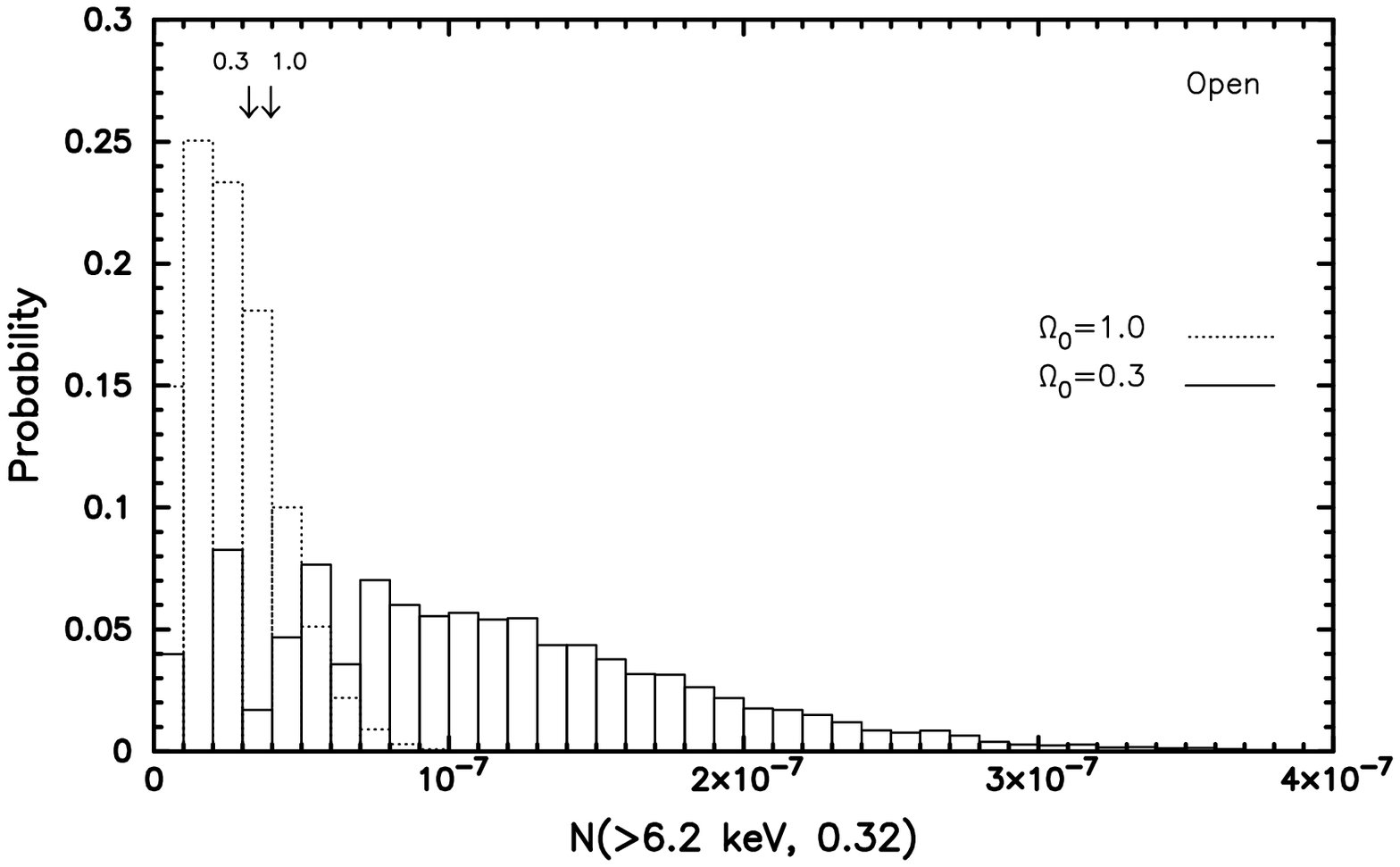}\\
\leavevmode\epsfysize=5.4cm \epsfbox{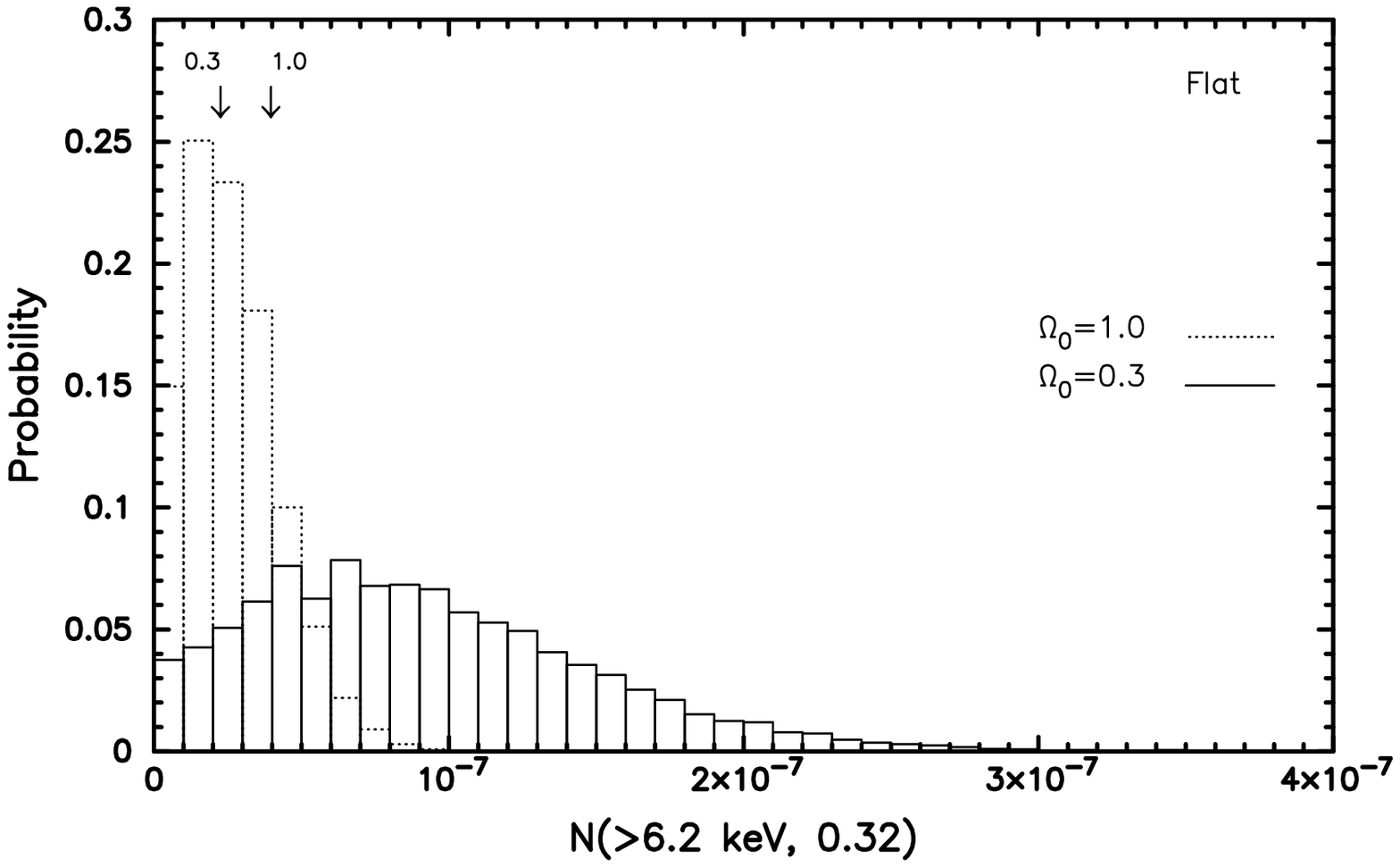}\\
\caption[Figure1]{Binned probability distributions for the expected value of
$N(>6.2\,{\rm keV}, 0.32)$ in the case of $\Omega_{0}=1$ (dotted lines) and
$\Omega_{0}=0.3$ (full lines).  They were obtained from $10^{4}$ realizations
of the bootstrap procedure described in the text.  Dotted lines correspond to
$\Omega_{0}=1$ and full lines to $\Omega_{0}=0.3$.  The arrows show the
corrected value for $N(>6.2\,{\rm keV}, 0.32)$ calculated using the dataset 
in Henry \shortcite{Henry} for both cases.  In the upper panel we assume the
cosmological constant to be zero, while in the lower panel we show the
results for a spatially-flat universe.}  
\end{figure}

The calculations we have just described assume that there is no scatter in
the relation between cluster X-ray temperature and luminosity.  This is not
correct, as mentioned previously, and can lead to an {\em increase} in the
estimated value for $N(>k_{\rm B}T,\,z)$.  This incompleteness problem
worsens as the threshold X-ray temperature $k_{\rm B}T$ is lowered, as one
starts considering clusters with X-ray fluxes dangerously close to the flux
detection limit.  For the same threshold X-ray temperature, the problem is
also potentially much more serious in the case of the $z=0.32$ data than for
the $z=0.05$ data.  The reason is simply that for the same flux detection
limit, the faintest clusters that can be detected nearby have X-ray
luminosities (and thus temperatures) which are considerably smaller than
those of the faintest clusters further away.

The effect of the scatter in the X-ray cluster temperature--luminosity
relation in the calculation of $N(>6.2\,{\rm keV},\,0.05)$ is negligible, as
the $z=0.05$ dataset in Henry \& Arnaud \shortcite{HA} is claimed to be
nearly complete down to at least 3 keV.

In the case of the $z=0.32$ data, it is not clear whether the presence of 
scatter in the X-ray cluster temperature--luminosity relation may affect 
the determination of $N(>6.2\,{\rm keV},\,0.32)$. Due to the scatter,
there is a finite probability that some of the 5 {\em EMSS}
galaxy clusters with X-ray flux below $2.5\times10^{-13}\;{\rm erg}\,{\rm
cm}^{-2}\,{\rm s}^{-1}$, that were found in the redshift range from 0.3 to 
0.4
\cite{Hetal}, may not only have an X-ray temperature in excess of the lowest
X-ray temperature present in the sub-sample of 10 clusters from Henry
\shortcite{Henry}, 3.8 keV for MS1512.4, but also in excess of our chosen
threshold temperature 6.2 keV.  The minimization of this possibility 
was in fact another reason for our choice of 6.2 keV as the threshold 
temperature. For example, at 4 keV the effect is already expected to be 
significant \cite{Ekeetal}.  

We calculated the expected increase in the {\em corrected} value of
$N(>6.2\,{\rm keV},\,0.32)$ given the dataset in Henry \shortcite{Henry}, as
a result of the existence of the 5 {\em EMSS} galaxy clusters mentioned
above, by doing 1000 Monte Carlo simulations where the X-ray temperatures for
those clusters were estimated via the X-ray cluster temperature--luminosity
relation determined in Eke et al.~\shortcite{Ekeetal} using the more recent 
data for the galaxy clusters in Henry \& Arnaud \shortcite{HA}.  The X-ray
temperatures for those 5 clusters were obtained through the power-law
relation
\begin{equation}
k_{\rm B}T=\left(10^{a}\,L_{44}^{X}\right)^{1/b}\,,
\end{equation}
where in each Monte Carlo simulation the values of the parameters $a$ and $b$ 
were drawn from Gaussian distributions with respectively mean 2.53 and 
dispersion 0.69, and mean 3.54 and dispersion 0.47. The X-ray luminosities 
in the 0.3 to 3.5 keV band for the 5 clusters in question in units of 
$10^{44}\,h^{-2}\,{\rm erg}\;{\rm s}^{-1}$, $L_{44}^{X}$, were determined 
using the X-ray fluxes given in Henry et al.~\shortcite{Hetal}, with 
a K-correction of 15 per cent included independently of the assumed cosmology 
\cite{Hetal,Henry}. The maximum volumes within which the 5 clusters 
could have been observed were calculated in the same way as those for the 10
higher flux clusters, using the {\em EMSS} data provided in Henry et 
al.~\shortcite{Hetal}. The final exclusion level for each value of 
$\Omega_{0}$ was calculated simply by taking the mean of the exclusion levels 
associated with the new higher estimated values for $N(>6.2\,{\rm 
keV},\,0.32)$ 
in the Universe, resulting from each Monte Carlo simulation. 

By using the X-ray cluster temperature--luminosity relation observed
for $z=0.05$ to estimate X-ray temperatures for galaxy clusters where
$z$ is within 0.3 to 0.4, we are implicitly assuming that this
relation does not evolve much between a redshift of 0.4 and the
present.  This is supported by the recent analyses of Mushotzky \&
Scharf \shortcite{MScharf} and Allen \& Fabian \shortcite{AF} (see
also Sadat, Blanchard and Oukbir 1998).  However, the actual dataset
in Henry \shortcite{Henry} does not support this assumption, as a
chi-square fit of a power-law to the data prefers the situation where
the X-ray luminosity is practically independent of the X-ray
temperature.  This is driven by the relatively low X-ray temperature
measured for the galaxy cluster MS2137.3, which, although it is by far
the brightest cluster in the dataset, has only the 7th highest X-ray
temperature.  Removing this cluster from the dataset, the best-fit
power-law for the X-ray temperature--luminosity relation becomes
compatible with the $z=0.05$ one.  It was this very strong dependence
of the best-fit X-ray cluster temperature--luminosity relation on the
inclusion or not of a single galaxy cluster which led us to choose not
to estimate the X-ray temperatures for the 5 clusters with the lowest
fluxes using the data for the 10 clusters that make up the dataset in
Henry \shortcite{Henry}.

In the end, we found that allowing for the presence of scatter in the X-ray
cluster temperature--luminosity relation when calculating $N(>6.2\,{\rm
keV},\,0.32)$ has only a small effect, at the few per cent level, 
on the exclusion levels obtained for different $\Omega_{0}$, and does 
not alter our conclusions. 

\section{Results}

The calculation of the theoretically-predicted overall value for 
$N(>6.2\,{\rm keV},\,0.32)$ was performed as 
described in Section 2, with $\sigma_{8}$ obtained from 
the observed value for $N(>6.2\,{\rm keV},\,0.05)$, and the 
uncertainty calculated via a Monte Carlo procedure. 

Our first result is the value required for $\sigma_{8}$, as a function of 
both $\Omega_{0}$ and $\lambda_{0}$, so that the observed value for 
$N(>6.2\,{\rm keV},0.05)$ is reproduced. This supersedes the result obtained 
in VL. We find that the best-fitting value is given by
\begin{displaymath}
\label{final1}
\sigma_8=\left\{ \begin{array}{ll}
0.56 \; \Omega_0^{-0.34} & {\rm Open}\\
0.56 \; \Omega_0^{-0.47} & {\rm Flat}.
\end{array}\right.
\end{displaymath}
This is accurate within 3 per cent for $\Omega_{0}$ between 0.1 and 1. 
 
The most important reason why this value is smaller than that quoted in VL 
is the decrease in the assumed number density of galaxy clusters at $z=0.05$. 
This results from the revision of the Henry \& Arnaud dataset and from 
the correction due to the existence of X-ray temperature measurement 
errors, which had not been taken into consideration in VL. Also, the 
cluster X-ray temperature function obtained in Henry \& Arnaud \shortcite{HA} 
had been slightly overestimated due to a calculational error \cite{ECF}. 

The overall uncertainty in the value of $\sigma_{8}$ was calculated in the 
same way as in VL, through a Monte Carlo procedure where the sources of 
error, namely the normalization of the cluster X-ray temperature to virial 
mass relation, the value of $\delta_{{\rm c}}$ and the value of $f$, are 
modelled as being Gaussian distributed, and $\Gamma$ and 
$N(>6.2\,{\rm keV}, 0.05)$ as having a lognormal distribution. As in VL, 
we find that for each $\Omega_{0}$ between 0.1 and 1 the distribution of 
$\sigma_{8}$ can be approximated by a lognormal. For open models, the 
95 per cent confidence limits are roughly given by $+20\Omega_0^{0.1{\rm 
log}_{10} \Omega_0}$ per cent and $-18\Omega_0^{0.1{\rm log}_{10} 
\Omega_0}$ per cent, while for flat models we have $+20\Omega_0^{0.2{\rm 
log}_{10} \Omega_0}$ per cent and $-18\Omega_0^{0.2{\rm log}_{10} \Omega_0}$. 

The calculation of the uncertainty in the theoretically-predicted 
overall value for $N(>6.2\,{\rm keV},\,0.32)$, for each assumed $\Omega_{0}$, 
was made using the Monte Carlo simulations performed in order to calculate 
the uncertainty in the value of $\sigma_{8}$, described above. 

We find that for $\Omega_{0}$ between 0.1 and 1, the distribution of 
$N(>6.2\,{\rm keV},0.32)$ is close to lognormal and, with an associated 
error of less than 4 per cent, its mean is fitted by 
\begin{equation}
\label{ndth}
N(>6.2\,{\rm keV},\,0.32)=2.67\Omega_{0}^{-G(\Omega_{0})}\times10^{-8}
h^{3}\;{\rm Mpc}^{-3}\,,
\end{equation}
where $G(\Omega_{0})=0.57 + 2.69\,\Omega_{0} - 1.87\,\Omega_{0}^{2}$ if the 
Universe is assumed open and $G(\Omega_{0}) = 0.45 + 2.56\,\Omega_{0} -
2.12\,\Omega_{0}^{2}$ if it is assumed spatially-flat.  The 95 per cent
confidence intervals are to a fair approximation given by
$+170\,\Omega_{0}^{0.17+0.31\Omega_{0}}$ per cent and 
$-62\,\Omega_{0}^{0.07+0.09\Omega_{0}}$ per cent in
the open case, and $+170\,\Omega_{0}^{0.06+0.42\Omega_{0}}$ per cent and
$-64\,\Omega_{0}^{0.03+0.16\Omega_{0}}$ per cent in the flat case.

As described in subsection 3.3, we can now build the distribution function 
one would expect to recover if $\Omega_{0}$ took a certain value in the 
Universe and $N(>6.2\,{\rm keV},\,0.32)$ was measured a large number of times 
across the sky under the same type of sampling that led to the dataset in 
Henry \shortcite{Henry}. Its mean is the theoretically-expected overall 
value for $N(>6.2\,{\rm keV},\,0.32)$ given in expression (\ref{ndth}) for 
the $\Omega_{0}$ under consideration, and the shape of the distribution 
is that obtained through the bootstrap procedure described in 
subsections 3.2 and 3.3. The exclusion level on the assumed $\Omega_{0}$ 
is then given by one minus the probability of measuring 
a value for $N(>6.2\,{\rm keV},\,0.32)$ as high (or as low) as that
implied by the dataset in Henry \shortcite{Henry}, calculated in 
expression (\ref{ndhzobs}), given the expected distribution. 

However, due to the uncertainties in the estimation of the
theoretically-expected overall value for $N(>6.2\,{\rm keV},\,0.32)$,
the actual calculation of the exclusion level for each $\Omega_{0}$ is
not as simple. So that we can obtain it, we need to integrate over all
possible values for the theoretically-expected overall $N(>6.2\,{\rm
keV},\,0.32)$, which we will denote $u$. The overall exclusion is the
product of the probability, $P(u,\Omega_{0})$, of each $u$ being
the correct overall value one would expect for $N(>6.2\,{\rm
keV},\,0.32)$ in the Universe (given the assumed $\Omega_{0}$), and the
exclusion level ${\rm Ex}(u)$ calculated as described above for each
assumed $u$, i.e.
\begin{equation}
\label{excl}
{\rm Exclusion}\,{\rm probability}\,{\rm of}\,\Omega_{0}=
\int_{-\infty}^{+\infty}P(u,\Omega_{0})\,{\rm Ex}(u)du
\end{equation}
As mentioned above, the $P(u,\Omega_{0})$ are lognormal distributions with 
mean given by expression (\ref{ndth}). The dispersion can be 
calculated from the 95 per cent confidence limits. 

In Figure~2 we show the exclusion levels for $\Omega_{0}$ obtained in
this way.  Even for the values of $\Omega_{0}$ for which it is easiest
to reproduce the observations, from 0.7 to about 0.8, the exclusion
level is quite high, around 70 per cent.  The reason lies with the
large uncertainty in the theoretically-expected overall value for
$N(>6.2\,{\rm keV},\,0.32)$.  Because of it, most
theoretically-expected overall values end up far away from the value
for $N(>6.2\,{\rm keV},\,0.32)$ which is expected in the Universe
given the dataset in Henry \shortcite{Henry}. A large uncertainty in
the theoretical prediction is clearly no basis to discard
models. However, for the high and low $\Omega_{0}$ we are aiming to
constrain, this effect becomes much less important; the high exclusion
levels are caused by most of the distribution for the
theoretically-expected overall values for $N(>6.2\,{\rm keV},\,0.32)$
being higher (for low $\Omega_0$), or lower (for high $\Omega_0$),
than the observations. Note that the exclusion levels are absolute,
and not relative as one would obtain from the calculation of a
likelihood function.

\begin{figure}
\centering
\leavevmode\epsfysize=5.4cm \epsfbox{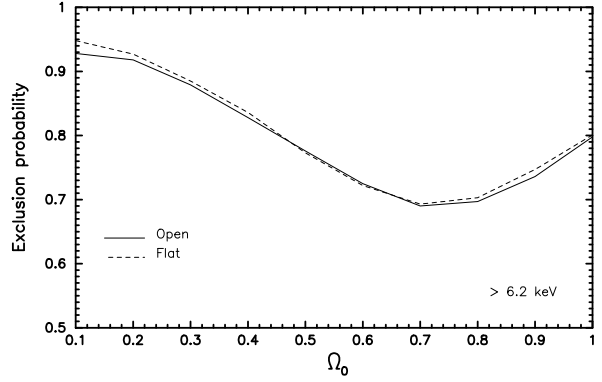}\\
\caption[Figure2]{The absolute exclusion levels for 
different values of $\Omega_{0}$ in both the open and 
spatially-flat cases, when the threshold X-ray temperature of
6.2 keV is used.}
\end{figure}

In the calculation of the theoretically-expected overall value for 
$N(>6.2\,{\rm keV},\,z=0.32)$, the parameter $f$ represents the fraction of 
the cluster mass at the redshift of cluster observation, $z_{\rm obs}$, 
that had been assembled by the time the cluster virialized, $z_{{\rm c}}$.  
We considered this parameter to be equal to 0.75, though we allowed for the 
possibility that it could be as low as 0.60 or as high as 0.90, 
corresponding this to something like a 95 per cent confidence interval. 
We modelled this uncertainty by assuming in the Monte Carlo simulations 
that $f$ was Gaussian distributed with mean 0.75 and dispersion 0.075. 

In order to estimate the effect of changing the assumed value for $f$ in the 
determination of $N(>6.2\,{\rm keV},\,z=0.32)$, we also performed 
calculations where we treated $f$ as having no associated uncertainty. 
We considered the cases where $f$ was equal either to 0.60, 0.75 or 0.90.  
We found that changing $f$ from 0.75 to 0.60 decreased the estimated 
value for $N(>6.2\,{\rm keV},\,z=0.32)$ by about 10 per cent, while changing 
$f$ from 0.75 to 0.90 increased $N(>6.2\,{\rm keV},\,z=0.32)$ by around 11 
per cent. The impact of removing the uncertainty in the value of $f$ from 
the estimation of the overall uncertainty in $N(>6.2\,{\rm keV},\,z=0.32)$ 
is therefore negligible. 

In all previous uses of the Press--Schechter framework to calculate
the evolution of the number density of rich galaxy clusters with
redshift \cite{OB,ECF,OB97,Henry,Ekeetal,Mark98,Retal}, it has been
assumed that the redshift of cluster virialization, $z_{{\rm c}}$,
coincides with that at which the galaxy cluster is observed, $z_{\rm
obs}$.  In Figure 3 we compare the value of $N(>6.2\,{\rm keV},\,z)$
obtained using the Lacey \& Cole (1993, 1994) prescription for the
estimation of $z_{{\rm c}}$ with the result of the assumption that $z_{{\rm
c}}=z_{{\rm obs}}$.  We always require that the observed value for
$N(>6.2\,{\rm keV},\,0.05)$ is recovered.

\begin{figure}
\centering
\leavevmode\epsfysize=5.4cm \epsfbox{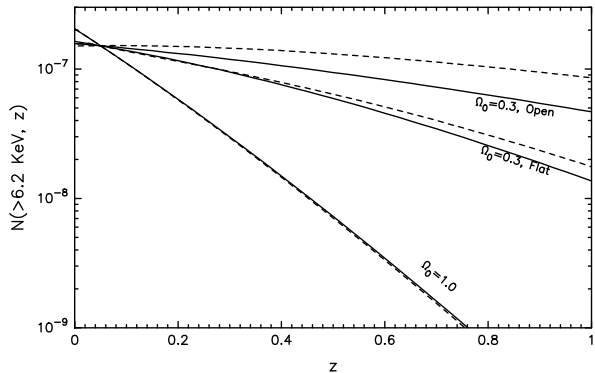}\\
\caption[Figure3]{The expected redshift evolution of 
$N(>6.2\,{\rm keV},\,z)$ for $\Omega_{0}=1$ and 0.3. The solid lines show the 
result obtained using the Lacey \& Cole method for estimating $z_{{\rm c}}$, 
and the dashed ones the result obtained assuming that $z_{{\rm c}}=z_{{\rm 
obs}}$. Each curve is normalized to reproduce the observed value for 
$N(>6.2\,{\rm keV},\,0.05)$. Note that the divergence at high $z$ is caused 
by this renormalization; the absolute correction is largest at the lowest 
redshift, where $\Omega(z)$ is smallest.}
\end{figure}

As expected, the difference in the theoretically-predicted overall 
value of $N(>6.2\,{\rm keV},z)$ 
resulting from the two distinct assumptions regarding $z_{{\rm c}}$ becomes
larger for $\Omega_{0}=0.3$, reflecting the fact that as $\Omega_{0}$ goes
down galaxy clusters tend to form increasingly at an earlier epoch than that
at which they are observed.  We find that neglecting the fact that some
clusters of galaxies virialize prior to the epoch at which they are observed
leads to an underestimation of the predicted degree of evolution
in the value of $N(>k_{\rm B}T,\,z)$ for $z>z_{\rm norm}$, where $z_{\rm
norm}$ is the redshift at which $N(>k_{\rm B}T,z)$ is normalized through
observations, e.g in our case $z_{\rm norm}=0.05$.  Taking into account the
possibility that $z_{{\rm c}}$ may be larger than $z_{\rm obs}$ therefore
requires {\em lower} values for $\Omega_{0}$ in order for the high-redshift
data on $N(>k_{\rm B}T,\,z)$ to be reproduced.

Allowing for $z_{{\rm c}}>z_{{\rm obs}}$ means that some galaxy clusters that
otherwise would not be massive enough to reach a given threshold temperature
$k_{\rm B}T$ can now be counted when calculating $N(>k_{\rm B}T,\,z)$.  In
principle this would have the effect of increasing the expected value of
$N(>k_{\rm B}T,\,z)$ for any $z$.  However, at the normalization redshift
0.05 the higher value for $N(>k_{\rm B}T,\,0.05)$ means that a less well
developed density field at $z=0.05$ is required, i.e.~a lower value of
$\sigma_{8}$ results from introducing the possibility that $z_{{\rm
c}}>z_{\rm obs}$.  As the number density of virialized objects evolves faster
for the same relative change in the value of the dispersion of the density
field the smaller this value is, the decrease in the required value for
$\sigma_{8}$ has the effect of enhancing the decrease in the value of
$N(>k_{\rm B}T,\,z)$ as $z$ gets larger.  This effect turns out to be more
important than the expected increase in the value of $N(>k_{\rm B}T,\,z)$ due
to higher cluster X-ray temperatures at fixed cluster mass resulting from the
possibility of $z_{{\rm c}}>z_{\rm obs}$.

\section{Discussion}

{}From the above analysis, we conclude that {\em at present} it is not
possible to reliably exclude any interesting value for $\Omega_{0}$ on
the basis of X-ray cluster number density evolution alone, due to the
limited statistical significance of the available observational data
and to uncertainties in the theoretical modelling of cluster formation
and evolution.  However, we do find that values of $\Omega_{0}$ below
0.3 are excluded at least at the 90 per cent confidence level.  Values
of $\Omega_{0}$ between 0.7 to 0.8 are those most favoured, though not
strongly.  These results are basically independent of the presence or
not of a cosmological constant.

Our conclusions support those of Colafrancesco, Mazzotta \& Vittorio 
\shortcite{CMV}, who tried to estimate the uncertainty involved in the 
estimation of the cluster X-ray temperature distribution function at 
different redshifts based on its present-day value. They found this 
uncertainty, given the still relatively poor quality of the data, to be 
sufficiently large to preclude the imposition of reliable limits on the 
value of $\Omega_{0}$. 
 
Our results disagree with those of Henry \shortcite{Henry} and 
Eke et al.~\shortcite{Ekeetal}, as they found the preferred $\Omega_{0}$ 
to lie between 0.4 to 0.5, with the $\Omega_{0}=1$ hypothesis strongly 
excluded. This disagreement is mainly the consequence of our focus on the 
threshold X-ray temperature of 6.2 keV, while they draw their conclusions 
based on the analysis of the results obtained for several threshold X-ray 
temperatures. Further below we will repeat our calculations assuming 
a threshold X-ray temperature of 4.8 keV, and we will find that 
when we calculate the joint probability of some value for 
$\Omega_{0}$ being excluded on the basis of the results concerning 
either one or both threshold X-ray temperatures of 6.2 keV and 4.8 keV,
the favoured value for $\Omega_{0}$ decreases to around 0.55. Some of the 
reasons for our choice of deriving the conclusions solely based on the 
results obtained for the 6.2 keV threshold were mentioned at the end of 
subsection 3.1 and others will be detailed below. 

Other less important contributions to the difference between our results 
and those presented by Henry \shortcite{Henry} and 
Eke et al.~\shortcite{Ekeetal} are the different assumed 
normalization for the virial mass to X-ray temperature relation, 
and the corrections in the expected values in the Universe for 
both $N(>6.2\,{\rm keV},\,0.05)$ and $N(>6.2\,{\rm keV},\,0.32)$ due to 
the uncertainties in the X-ray cluster temperature measurements. 
Note that changing the mean of the bootstrap distribution 
obtained for $N(>6.2\,{\rm keV},\,0.32)$ to its 
theoretically-expected overall value in some $\Omega_{0}$ universe
and then calculating the exclusion level on the estimated value for 
$N(>6.2\,{\rm keV},\,0.32)$ in the Universe given the dataset 
in Henry \shortcite{Henry}, rather than just using the original 
bootstrap distribution to impose an exclusion level on the 
theoretically-expected overall value for $N(>6.2\,{\rm keV},\,0.32)$ 
in that $\Omega_{0}$ universe, does not seem to make much difference. 
This is a reflection of the fact that the bootstrap distributions 
recovered do not have a strongly asymmetric shape. 

Our disagreement with Eke et al.~\shortcite{Ekeetal} on the level of 
exclusion of the $\Omega_{0}=1$ hypothesis is also due to our much larger 
assumed uncertainty in the theoretically-expected overall value for 
$N(>6.2\,{\rm keV},\,0.32)$. 

For the $\Omega_{0}=1$ hypothesis to be favoured, one requires the
lowest possible observed value for $N(>6.2\,{\rm keV},0.32)$. This is
best achieved if, for the sample of 10 galaxy clusters used in its
calculation, the X-ray temperatures turn out to be on average lower
than the assumed mean, and the X-ray fluxes higher. A higher ratio
between the extended and detect cell fluxes for the {\em EMSS} at
$z=0.32$ would also help.  On the theoretical side, the higher one
decides the expected value for $N(>6.2\,{\rm keV},0.32)$ is, the more
compatible with the data the $\Omega_{0}=1$ hypothesis becomes.  This
can be best achieved if, in decreasing order of importance, the
cluster virial mass at fixed X-ray temperature is being
underestimated, $\delta_{{\rm c}}$ is lower than the canonical value
1.7 and $f$, the assembled fraction of a cluster virial mass after
which the X-ray temperature does not change significantly, is assumed
greater than 0.75.  However, the single most important factor in
determining the theoretically-expected overall value for $N(>6.2\,{\rm
keV},0.32)$ is the present-day normalization for the dispersion of the
density field, $\sigma_{8}$, which in turn results from
the observational value for the present density $N(>6.2\,{\rm
keV},\,0.05)$.

Although we worked with all X-ray clusters that make up the dataset in
Henry \shortcite{Henry}, and even estimated the effect of also
considering the 5 clusters with lower X-ray fluxes present in the {\em
EMSS} in the redshift bin from 0.3 to 0.4, in fact we only used the
abundance of clusters with X-ray temperatures in excess of 6.2 keV to
constrain $\Omega_{0}$. We mentioned some of the reasons for this
choice in Section~3. Nevertheless, we decided to repeat the same
calculations for a threshold X-ray temperature of 4.8 keV.
This value also well represents the mean curve going through the
observed cumulative X-ray temperature distribution function at both
$z=0.05$ and $z=0.32$.

\begin{figure}
\centering
\leavevmode\epsfysize=5.4cm \epsfbox{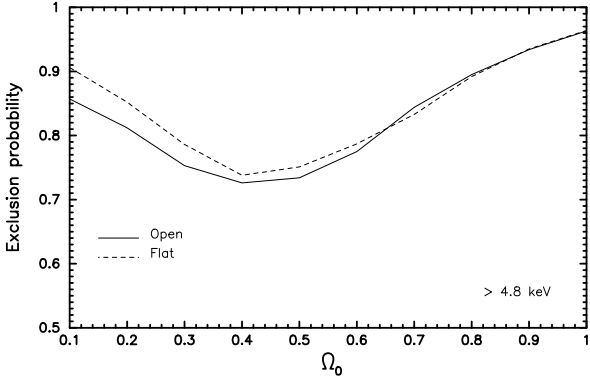}\\
\caption[Figure4]{The absolute exclusion levels for 
different values of $\Omega_{0}$ in both the open and 
spatially-flat cases, when the threshold X-ray temperature of
4.8 keV is used.}
\end{figure}

The results regarding the best-fit value for $\Omega_{0}$, presented in
Figure~4, are somewhat different from those we obtained when the threshold
X-ray temperature was assumed to be 6.2 keV.  This is particularly true if
the correction for the possibility of any of the 5 clusters with the lowest
X-ray fluxes in the $0.3<z<0.4$ EMSS sub-sample having X-ray temperatures in
excess of 4.8 keV is included, as can be seen in Figure 5 for the open case.
While the standard analysis without these 5 X-ray clusters prefers a value
for $\Omega_{0}$ between 0.4 to 0.5, when the correction for the scatter in
the relation between the cluster X-ray temperature and luminosity is
included, in the way described in subsection 3.3, the preferred value for
$\Omega_{0}$ decreases to about 0.3.  Now the $\Omega_{0}=1$ hypothesis is
excluded at more than the 95 per cent confidence level, with or without the
correction.  At the 90 per cent confidence level, one finds that
$\Omega_{0}>0.8$ is excluded without the correction, being this limit lowered
to 0.7 when the correction is included. 

One can also estimate the joint probability of some $\Omega_{0}$ value being 
excluded on the basis of the results relative to either one or both  
X-ray temperature thresholds. Assuming the data used in the 
calculations for the two thresholds is independent, the results then 
imply that the favoured value for $\Omega_{0}$ is close to 0.55 
(0.50 if the incompleteness correction is included) and the 
$\Omega_{0}=1$ hypothesis is excluded at the 99 per cent level. This agrees 
very well with the results of Henry \shortcite{Henry} and 
Eke et al.~\shortcite{Ekeetal}, leading us to believe that the main 
difference between our analysis and theirs is our decision to 
draw our conclusions solely based on the exclusion levels obtained for the 
X-ray temperature threshold of 6.2 keV. 

A further potential problem one must consider when working with
clusters whose observed X-ray temperature is as low as 4.8 keV is the
possibility that the energy in the intracluster gas has increased as a
result of (pre-)heating by supernovae and starbursts in the cluster
galaxies.  In fact this is the leading hypothesis (e.g.  Navarro,
Frenk \& White 1995; Markevitch 1998) put forward to explain the
discrepancy between the observed slope of the X-ray
temperature--luminosity relation, close to 0.3, and the expected value
of 0.5 if clusters evolve in a self-similar way \cite{Kaiser}.

Following Eke et al.~\shortcite{Ekeetal}, we assume that in a cluster
whose observed X-ray temperature is 4.8 keV, 17 per cent of its
energy, that is 0.8 keV per intracluster gas particle, was due to
(pre-)heating produced by processes occurring inside the cluster
galaxies. This is approximately the amount of energy that gets
injected into the intracluster gas particles in the simulation of
Metzler \& Evrard \shortcite{ME}, where a galaxy cluster's X-ray
temperature, which would otherwise be 5.6 keV, increased to 6.4 keV.
Note however that in the scheme proposed by Eke et
al.~\shortcite{Ekeetal} a cluster this large would not be (pre-)heated
to the extent simulated by Metzler \& Evrard \shortcite{ME}, as in
their proposal Eke et al.~\shortcite{Ekeetal} assume that the energy
gained by each intracluster gas particle due to (pre-)heating
decreases as a galaxy cluster becomes larger, being close to zero for
galaxy clusters with X-ray temperatures exceeding 6.2 keV.

\begin{figure}
\centering
\leavevmode\epsfysize=5.4cm \epsfbox{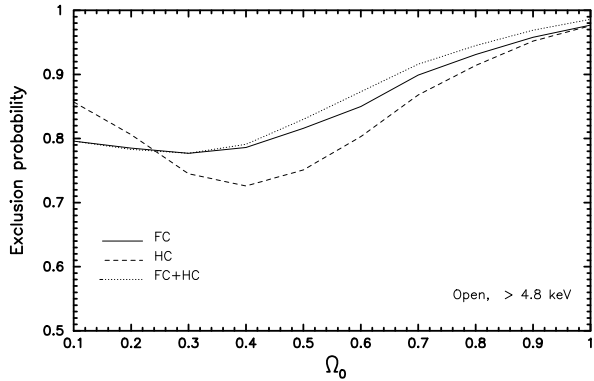}\\
\caption[Figure5]{The absolute exclusion levels for 
different values of $\Omega_{0}$ only for the open case, 
when the threshold X-ray temperature of 4.8 keV is used. 
The full curve includes a correction (FC) for the possibility of the 
5 clusters with lowest fluxes in the {\em EMSS} located between 
$z=0.3$ and $z=0.4$ having X-ray temperatures in excess of 4.8 keV. 
The dashed curve includes a correction (HC) for the possibility of 
(pre-)heating of the intracluster medium due to processes within the 
cluster galaxies. The dotted curve includes both corrections.}
\end{figure}

The above assumption means that the observed values for $N(>4.8\,{\rm
keV},\,z)$, when $z=0.05$ and $z=0.32$, should now be compared with
the theoretically-expected values for $N(>4.0\,{\rm keV},\,z)$ at
those redshifts.  The resulting exclusion levels on the value of
$\Omega_{0}$ can be seen in Figure 5 for the open case.  There is
little difference compared to the results in Figure 4 that follow from
the standard no-heating calculation.  The lower value for
$\sigma_{8}$, required to match theory and observations at $z=0.05$,
more than compensates for the expected increase in the number of
galaxy clusters with X-ray temperatures in excess of 4.8 keV at
$z=0.05$, in effect bringing this number down.  In fact, the standard
no-heating calculation for a threshold X-ray temperature of 4.8 keV
requires a value for $\sigma_{8}(\Omega_{0})$, so that the observed
value for $N(>4.8\,{\rm keV},\,0.05)$ is reproduced, that is less than
3 per cent below that required by the $>6.2$ keV data, quoted in
equation (\ref{final1}).  On the other hand, including the
(pre-)heating correction, the required $\sigma_{8}(\Omega_{0})$ value
drops to 19 per cent below that preferred by the $>6.2$ keV data.
Though the coincidence between the $\sigma_{8}$ values obtained for 
the two X-ray temperature thresholds 4.8 keV and 6.2 keV under 
the no-heating assumption may be accidental, it could indicate that
(pre-)heating was relatively unimportant at least for the galaxy
clusters observed at $z=0.05$ with X-ray temperatures exceeding 4 keV.
If (pre-)heating was more important in the past than today, then the
required $\sigma_{8}(\Omega_{0})$ value would be that obtained through
the standard no-heating hypothesis, but the comparison at $z=0.32$
would include the (pre-)heating correction.  This would push the
theoretically-expected value for $N(>4.8\,{\rm keV},\,0.32)$ up,
favouring higher values for $\Omega_{0}$.  This is not as far-fetched
as it may seem, given that it is well known that the star-formation
rate peaks before $z=1$ (e.g.  Madau, Ferguson \& Dickinson 1998;
Baugh et al.  1998), and consequently so does the rate of supernovae
Type II (the rate of supernovae Type Ia peaks a few Gyr later) and the
probability of starbursts.

The results for the 4.8 keV threshold X-ray temperature are close to
those found by Eke et al.~\shortcite{Ekeetal}, leading us to believe
that their exclusion levels for $\Omega_{0}$ are dominated by the
information associated with the threshold X-ray temperatures 4.0 keV
and 5.0 keV. In our view, the analysis for these X-ray temperature
thresholds carries with it a sufficient number of uncertainties, due
to the problems mentioned above, so as to render the constraints
imposed on $\Omega_{0}$ not very trustworthy.  Only the data regarding
clusters with X-ray temperatures in excess of about 6 keV seems
sufficiently free of modelling problems so as to be potentially useful
in constraining $\Omega_{0}$.

Another possible complication has arisen from recent work by 
Blanchard, Bartlett and Sadat \shortcite{BBS} who use a sample 
of 50 galaxy clusters with mean redshift of 0.05, which were identified 
through the {\em ROSAT} satellite, to estimate the cumulative X-ray 
temperature distribution function at $z=0.05$. They claim the number density 
of galaxy clusters at $z=0.05$ with X-ray temperatures exceeding 4 keV 
is being {\em underestimated} when the Henry \& Arnaud cluster sample is 
used. Through the X-ray cumulative temperature distribution function 
at $z=0.05$ they obtain, they then estimate $\Omega_{0}$ using the 
{\em EMSS} cluster abundance in the redshift bin $0.3<z<0.4$ and the X-ray 
temperature data gathered in Henry \shortcite{Henry}. They find 
the favoured value for $\Omega_{0}$ to be 0.75, while $\Omega_{0}<0.3$ is 
excluded at more than the 95 per cent level. These results coincide 
very well with ours when only the 6.2 keV threshold X-ray temperature 
is considered, thus perhaps implying that the discrepancy between the 
favoured 
value for $\Omega_{0}$ found when different X-ray temperature thresholds are 
considered may arise from a underestimation of the cumulative distribution 
function at $z=0.05$ for X-ray temperatures below about 6 keV. 

Unfortunately, due to uncertainties associated both with the observational 
measurements and the theoretical modelling of cluster evolution, 
the presently-available data on galaxy clusters 
with X-ray temperatures exceeding about 6 keV is not able to strongly
discriminate between cosmologies with different values for $\Omega_{0}$.  
And in any case, the data available is probably not yet statistically
significant.  More is needed to support or disclaim the preliminary
conclusions that can be obtained from it.  In particular there are some
oddities with the sub-sample of {\em EMSS} galaxy clusters observed by
Henry, such as the strange redshift distribution, strongly clustered
around $0.32$, and the unexpectedly low X-ray temperature of MS2137.3, that
makes one have some doubts about how representative this dataset is of the 
Universe.

Within the next few years, with the launch of the {\em XMM} satellite,  
possibly in late 1999, a significant increase in the quantity and quality 
of the available data is expected to occur \cite{Romer}. It should then be 
possible to place stronger constraints on $\Omega_{0}$ 
on the basis of the evolution of the galaxy cluster X-ray temperature 
function. This would be helped by improvements in the theoretical 
modelling of cluster evolution, perhaps based on the high-resolution 
hydrodynamical $N$-body simulations on cosmological scales expected 
in the near future.

\section*{ACKNOWLEDGMENTS}

We are very grateful to Patrick Henry for his help in clarifying the 
procedure leading to the calculation of the detection volumes, and for 
providing us with the results of his recalculation of the present-day 
integral cluster X-ray temperature distribution. We also thank Alain 
Blanchard, Lauro Moscardini and the referee, Vincent Eke, for many useful 
comments and discussions. PTPV is supported 
by the PRAXIS XXI program of FCT (Portugal), and ARL by the Royal Society. 



\bsp

\begin{thebibliography}{}
\bibitem[\protect\citename{Allen \& Fabian }1998]{AF} Allen S. W., 
        Fabian A. C., 1998, MNRAS, 297, L57
\bibitem[\protect\citename{Baugh et al. }1998]{BCFL} Baugh C.M., Cole S., 
        Frenk C.S., Lacey C.G., 1998,  ApJ, 498, 504
\bibitem[\protect\citename{Bernardeau }1994]{Ber} Bernardeau F.,
        1994, ApJ, 427, 51
\bibitem[\protect\citename{Blanchard \& Bartlett }1997]{BB} Blanchard A.,
        Bartlett J. G., 1997, A\&A, 332, L49
\bibitem[\protect\citename{Blanchard, Bartlett \& Sadat }1998]{BBS} Blanchard 
A.,
        Bartlett J. G., Sadat R., 1998, to appear in Les Comptes Rendus 
	de l'Academie des Sciences
\bibitem[\protect\citename{Bond et al. }1991]{BCEK} Bond J. R., Cole S.,
        Efstathiou G., Kaiser N., 1991, ApJ, 379, 440
\bibitem[\protect\citename{Bryan \& Norman }1998]{BN} Bryan G. L., 
        Norman M. L., 1998, ApJ, 495, 80
\bibitem[\protect\citename{Colafrancesco, Mazzotta \& Vittorio }1997]{CMV} 
        Colafrancesco S., Mazzotta P., Vittorio N., 1997, ApJ, 488, 566 
\bibitem[\protect\citename{David et al. }1993]{Davetal} David L. P., Slyz
        A., Jones C., Forman W., Vrtilek S. D., Arnaud K. A., 
        1993, ApJ, 412, 479
\bibitem[\protect\citename{Edge et al. }1990]{ESFA} Edge A. C., Stewart G.
        C., Fabian A. C., Arnaud K. A., 1990, MNRAS, 245, 559
\bibitem[\protect\citename{Eke et al. }1996]{ECF}
        Eke V. R., Cole S., Frenk C. S., 1996, MNRAS, 282, 263
\bibitem[\protect\citename{Eke et al. }1998]{Ekeetal} Eke V. R., Cole S.,
        Frenk C. S., Henry P., 1998, IoA preprint astro-ph/9802350
\bibitem[\protect\citename{Evrard }1989]{Evrard} Evrard A. E., 1989, 
        ApJ, 341, L71
\bibitem[\protect\citename{Fabian et al. }1994]{Fetal} Fabian A. C., 
        Crawford C. S., Edge A. C., Mushotzky R. F., 1994, MNRAS, 267, 779
\bibitem[\protect\citename{Fan, Bahcall \& Cen }1997]{FBC} Fan X., 
        Bahcall N. A., Cen R., 1997, ApJ 490, L123
\bibitem[\protect\citename{Frenk et al. }1990]{FWED} Frenk C. S.,
        White S. D. M., Efstathiou G., David M., 1990, ApJ, 351, 10
\bibitem[\protect\citename{Gioia \& Luppino }1994]{GioiaL} Gioia I. M., 
        Luppino G. A., 1994, ApJS, 94, 583
\bibitem[\protect\citename{Gioia et al. }1990]{Getal} Gioia I. M., Henry 
        J. P., Maccacaro T., Morris S. L., Stocke J. T., Wolter A.,
        1990, ApJ, 356, L35
\bibitem[\protect\citename{Gross et al. }1997]{Grossetal} Gross M. A. K.,
        Somerville R. S., Primack J. R., Borgani S, Girardi M., 1997,
        Santa Cruz preprint astro-ph/9711035
\bibitem[\protect\citename{Hanami }1993]{H} Hanami H., 1993, ApJ, 415, 42
\bibitem[\protect\citename{Hattori \& Matsuzawa }1995]{HM} Hattori M., 
        Matsuzawa H., 1995, A\&A 300, 637
\bibitem[\protect\citename{Henry }1997]{Henry} Henry J. P., 
        1997, ApJ, 489, L1
\bibitem[\protect\citename{Henry \& Arnaud }1991]{HA} Henry J. P., 
        Arnaud K. A., 1991, ApJ, 372, 410
\bibitem[\protect\citename{Henry et al. }1992]{Hetal} Henry J. P., Gioia
        I. M., Maccacaro T., Morris S. L., Stocke J. T., Wolter A., 
        1992, ApJ, 386, 408
\bibitem[\protect\citename{Kaiser }1986]{Kaiser} Kaiser N., 1986, 
         MNRAS, 222, 323
\bibitem[\protect\citename{Kitayama \& Suto }1997]{KSuto} Kitayama T., 
        Suto Y., 1997, ApJ, 490, 557
\bibitem[\protect\citename{Lacey \& Cole }1993]{LC93} Lacey C., Cole S.,
        1993, MNRAS, 262, 627
\bibitem[\protect\citename{Lacey \& Cole }1994]{LC} Lacey C., Cole S., 1994,
        MNRAS, 271, 676
\bibitem[\protect\citename{Lilje }1992]{lilje} Lilje P. B., 1992, ApJ,
        386, L33
\bibitem[\protect\citename{Madau, Ferguson \& Dickinson }1998]{Madau} Madau 
P., 
	Ferguson H.C., Dickinson M., 1998, ApJ, 498, 106
\bibitem[\protect\citename{Markevitch }1998]{Mark98} Markevitch M., 1998, 
        CfA preprint astro-ph/9802059
\bibitem[\protect\citename{Metzler \& Evrard }1994]{ME} Metzler C. A., 
        Evrard A. E., 1994, ApJ, 437, 564
\bibitem[\protect\citename{Monaco }1995]{Mon} Monaco P., 1995, 
        ApJ, 447, 23
\bibitem[\protect\citename{Mushotzky \& Scharf }1997]{MScharf} Mushotzky 
        R. F., Scharf C. A., 1997, ApJ, 482, L13
\bibitem[\protect\citename{Navarro et al. }1995]{NFW} Navarro J. 
        F., Frenk C. S., White S. D. M., 1995, MNRAS, 275, 720
\bibitem[\protect\citename{Nichol et al. }1997]{Netal} Nichol R. C., 
        Holden B. P., Romer A. K., Ulmer M. P., Burke D. J., Collins
        C. A., 1997, ApJ, 481, 644
\bibitem[\protect\citename{Norman \& Bryan }1998]{NB} Norman M. L., 
        Bryan G. L., 1998, to appear in the proceedings of the Ringberg
        Workshop on M87, eds. Meisenheimer K., Roeser H-J., Springer 
        Verlag, astro-ph/9802335
\bibitem[\protect\citename{Oukbir, Bartlett \& Blanchard }1997]{OBB} 
        Oukbir J., Bartlett J. G., Blanchard A., 1997, A\&A, 320, 365 
\bibitem[\protect\citename{Oukbir \& Blanchard }1992]{OB} Oukbir J.,
        Blanchard A., 1992, A\&A, 262, L21 
\bibitem[\protect\citename{Oukbir \& Blanchard }1997]{OB97} Oukbir J.,
        Blanchard A., 1997, A\&A, 317, 1
\bibitem[\protect\citename{Peacock }1997]{Pea} Peacock J. A., 
        1997, MNRAS, 284, 85
\bibitem[\protect\citename{Peacock \& Dodds }1994]{PD} Peacock J. A., Dodds 
        S. J., 1994, MNRAS, 267, 1020
\bibitem[\protect\citename{Press \& Schechter }1974]{PS} Press W. H.,
        Schechter P., 1974, ApJ, 187, 452
\bibitem[\protect\citename{Press et al. }1992]{PTVF} Press W. H., 
	Teukolsky S. A., Vetterling W. T., Flannery B. P., 1992, 
	{\em Numerical Recipes}, 2nd edition, Cambridge University Press
\bibitem[\protect\citename{Reichart et al. }1998]{Retal} Reichart D. E., 
        Nichol R. C., Castander F. J., Burke D. J., Romer A. K.,
        Holden B. P., Collins C. A., Ulmer M. P., 1998, 
        Chicago preprint astro-ph/9802153
\bibitem[\protect\citename{Romer }1998]{Romer} Romer A. M., 1998, 
	to appear in the proceedings of the 14th IAP meeting, 
	``Wide Field Surveys in Cosmology'', eds. Mellier Y., Colombi S., 
	astro-ph/9809198
\bibitem[\protect\citename{Sadat, Blanchard \& Oukbir }1998]{SBO} Sadat R., 
        Blanchard A., Oukbir J., 1998, A\&A, 329, 21
\bibitem[\protect\citename{Tormen }1998]{T} Tormen G., 1998, MNRAS, 297, 648
\bibitem[\protect\citename{Viana \& Liddle }1996]{VL} Viana P. T. P., 
        Liddle A. R., 1996, MNRAS, 281, 323 [VL]
\bibitem[\protect\citename{White et al. }1993a]{WEF} White 
        S. D. M., Efstathiou G., Frenk C. S., 1993a, MNRAS, 262, 1023
\bibitem[\protect\citename{White et al. }1993b]{WNEF} White S. D. M.,
        Navarro J. F., Evrard A. E., Frenk C. S., 1993b, Nat, 366, 429
\end{thebibliography}
\end{document}